# Influence of Filamentary and Strand Aspect Ratios on AC Loss in Short, Untwisted Samples of HTSC and LTSC Superconducting Multifilamentary Composites

M.D. Sumption, E. Lee, and E.W. Collings
LASM, MSE, The Ohio State University
Columbus, OH 43210

Oct 7, 1999

## Abstract

Measurements of magnetization, $M$, vs. magnetic field strength, $H$, have been made on HTSC Bi:2223/Ag and LTSC NbTi/matrix multifilamentary tapes as a function of ramp rate. The integrated loss, $Q_{tot}$, was extracted from these $M$-$H$ loops which had a maximum field amplitude of 17 kOe and ramp rates, $dH/dt$, ranging from 20 Oe/s to 700 Oe/s. All measurements were taken at 4.2 K, with $H$ directed normal and parallel, respectively, to the broad tape face. The samples were from 0.5 to 1.5 cm in length, and untwisted. The expressions needed to properly extract effective transverse matrix resistivities, $r_\wedge$, are discussed and applied. In general these expressions indicate that transverse loss in untwisted strands grows linearly with the aspect ratio, in contrast to that in twisted strands, whose losses grow with the square of the aspect ratio. With low aspect-ratio LTSC tapes our equations gave the expected values of $r_\wedge$. However, departures from the expected $r_\wedge$ emerged with further flattening of the strands as the filamentary array itself became sparse and highly aspected, and the filaments themselves became flattened. A model attributing the anomalous $r_\wedge$ to the effects of these features is proposed.

**Key words:** Bi:2223/Ag, flat multifilamentary tape, AC loss, effective matrix resistivity

## Introduction

The general expressions for AC loss in twisted multifilamentary composites is well known, having been calculated by Carr[1], and several others[2]. Carr's extensive work was a detailed examination of loss in mostly round strands, using an anisotropic-continuum-type calculation. He also calculated the loss for twisted aspected strands[3], as well as untwisted round strands[4]. With the advent of HTSC, Kwasnitza and Clerc adapted expressions for Rutherford-type cables to describe losses in flattened strands[5], obtaining results similar to those of Carr.

Although equipment for AC loss measurement in long samples is available, it is common for AC loss to be measured on short, straight samples, either untwisted or with just a few twists. The present paper focusses on this type of measurement. As for the presentation of AC loss results, while the loss data themselves are useful for strand-to-strand intercomparison, it is more meaningful and generally useful to parameterize loss quantitatively in terms of an intrinsic property of the strand, in particular its effective transverse matrix resistivity, $r_\wedge$. To enable a $r_\wedge$ to be calculated exact equations are needed. In what follows we discuss the present status of such equations as they apply to twisted and untwisted round and flattened strands and apply the appropriate ones to our experimental results on highly aspected untwisted strands. We then go on to demonstrate some significant departures from theory, and offer a model to explain them in terms of filamentary array- and filament aspect ratios, and filamentary array number density.

## Theory: Strand Aspect Ratio

Our analysis begins with the Carr "anisotropic-continuum" expression[1] for the AC loss in a *round* twisted MF strand exposed to a transverse sinusoidal magnetic field of amplitude $H_m$ and frequency $f$, viz:

$$Q_e = \frac{L_p^2}{2x10^9\, r_\perp} H_m^2 f \qquad (1)$$

in c.g.s units (or SI units with removal of the factor $10^9$). Here $Q_e$ is the eddy current coupling loss, which when added to the filamentary hysteretic loss gives the total loss, $Q_{tot}$ and $r_\wedge$ is the effective matrix resistivity which we seek in parameterizing our loss results. A somewhat different approach was adopted by Campbell[2], who found that for *twisted* aspected strands (aspect ratio $a/b$, sinusoidal $H$ perpendicular to $a$) $Q_e$ can be described by

$$Q_e = 2n\frac{p^2}{10^9} H_m^2 f\, t \qquad (2)$$

which clearly represents $Q_e$ as the product of an external shape factor, $n$ (which is a function of the demagnetization), and a relaxation factor , $t$. This single expression describes the $Q_e$ of several strand types: (i) *round twisted* with $n = 2$ and $t = L_p^2/r_\wedge(1/8p^2)$, (ii) $a \gg b$ *twisted* with $n = a/b$ and $t = L_p^2/r_\wedge(a/b)(7/480)$, (iii) $a \ll b$ *twisted* with $n = 1$ and $t = L_p^2/r_\wedge(a/b)^2(1/16)$.

Note that for the round twisted case, Campbell and Carr agree exactly (indicating the presence in Eqn. (1) of an *implicit n = 2* factor, as would be expected). Otherwise, apart from some more-or-less small prefactor differences, Campbell's expressions are similar to those of Carr (above) and Kwasnitza and Clerk[5] in the appropriate *a/b* limit. Corrected for wave shape (sinusoidal, triangular) they are generally valid for both flat strands and cables (e.g. Takács[6], flat cable). It is important to re-emphasize that the significance of the Campbell approach lies in the separability of $Q_e$ into contributions from (i) external-shape (via the demagnetization-associated shape factor, $n = 1/(1 - N)$) and (ii) internal structure. i.e. materials properties/strand design (via *t*, which embodies $r_\wedge$, $L_p$, and also *a/b*).

*Eddy current loss in untwisted strands:* Carr's basic expression for round untwisted strands[4] of length $L$ is the same as (1) except for a prefactor 0.81 and the replacement of $L_p$ by $2L$. With *n* no longer implicit the Carr expression becomes:

$$Q_e = \frac{0.81}{4} n \frac{L^2}{10^9 \, r_\perp} H_m (dH/dt) \qquad (3)$$

where a conversion from sinusoidal frequency to field ramp rate *dH/dt* was achieved using $4H_m f = dH/dt$. It can easily be shown that for untwisted strands of any aspect ratio the kernel of Eqn. (3) remains unchanged, which implies an untwisted *t* that is independent of *a/b*. Equation (3) thus represents an extension of Campbell's Eqn. (2) into the realm of untwisted strands.

Derived in this way, Eqn. (3) then represents the loss of an untwisted, aspected strand, treated as an anisotropic continuum. As such it should be directly applicable to flat densely filamented NbTi/CuMn tapes. We demonstrate that it is for additional clarity and to prove the correctness the expression. However, in order to treat the case of HTSC/Ag stapes, it is necessary to account for the unusual filamentary structure characteristic of these new materials, viz. filament aspect ratio and filament-array coarseness. The need for this is demonstrated by the results of experiments also performed on NbTi/Cu tapes, but his time with a relatively low number density of filaments and over a wide range of aspect ratios. The way that these factors influence enter into the volume-averaged $Q_e$ and hence $r_\wedge$ is described in terms of the following heuristic model.

**Theory: Filamentary Aspect Ratio and Array Structure**

While Eqn (3) accounts for the external aspect ratio of the strands, the internal configurations must be accounted for. In order to graphically demonstrate the influence that filamentary array structure and coarseness can have, consider the three strand configurations in *Figure* 1. All of these configurations have the same superconductor to total fraction, *l*. However, the distribution of the superconducting regions is quite different in the three cases. In the "z-stack" arrangement of *Figure* 1 (a) and for fields applied face-on (FO) to the strand (perpendicular to the wide side of the tape) the emf loops are generated and the resulting eddy currents flow in planes which are either entirely superconducting or entirely non-superconducting. This leads to the absence of any real "coupling" currents. The only significant loss is the hysteretic filamentary contribution.

In *Figures* 1 (b) and 1 (c), the superconducting regions are discontinuous and arranged in distinctly different patterns. In these cases, there are regions where emf loops cut across both superconducting and non-superconducting regions. This allows for the creations of eddy current coupling loops. Qualitatively, we expect for the losses to follow the standard expressions when the filament number densities are high, but for the **eddy current portion** of the losses to tend to zero as the filaments become larger and the array coarser. It must be noted at this point that such modifications are expected to be much weaker for twisted strands, since then the current paths tend to circulate on the outside of the filamentary array at low *dH/dt* values, only crossing the strand near its center[1]. However, it could be expected that for strands with resistive ribbons on the inside, the current paths will again resemble those of untwisted strands, and the loss reduction factors will be similar to those of untwisted strands.

## Experimental

*Strand Materials:* Three different types of multifilamentary superconducting materials were used: (a) Fine-array NbTi/CuMn strands (NB5000) (b) coarse-array NbTi/Cu strands (NB54), and (c) Bi:2223/Ag strands (BI) of various designs. These strand types and the conditions for which they were prepared for measurement are described in Tables 1-4.
NB5000 was an experimental NbTi strand which employed a CuMn matrix for proximity effect suppression and had a circular (fully filamentary) array, manufactured by Supercon. The NB5000 sample set was prepared at the Ohio State University (OSU) to several aspect ratios by rolling, pressing, or filing. The NB54 set was prepared by rolling at OSU. This entire series was vacuum annealed for 8h/400°C in order to remove the effects on the matrix resistivity of cold deformation. This precaution was not taken in preparing the NB5000 set since the deformation range was smaller and the matrix was a resistive alloy, Cu-0.6at.%Mn.

The Bi:2223/Ag sample set was prepared and delivered to us by the University of Wollongong. Sample BI19 was provided in both round and aspected forms, while BI36, BI36R, and BI9 were provided only as tapes.

*Measurement:* Measurements of magnetization vs. magnetic field strength, hence the area (total AC loss, $Q_{tot}$) and height $\Delta M(H)$ of the hysteresis loop, $M(H)$, were made using a vibrating sample magnetometer. All measurements were made at 4.2 K with the time-varying field, $H(t)$, applied either parallel (EO) or perpendicular (FO) to the strand's broad face  The waveform of $H(t)$ was triangular, with an amplitude ($H_m$) of 17 kOe, and a ramp rate, *dH/dt*, of  20 to 700 Oe/s. Additionally, four point resistivity ratio, RR, measurements were made on samples of BI36 and NB54. In the case of BI36, the measurements were made on a length of Ag sheath removed from a length of the original monocore after it had been annealed in air for 24h/800°C.

## Results

The Eq. (3) prediction has been confirmed for the NB5000 set. The dependence of $Q_e$ on aspect ratio ($n$, or $a/b$) is illustrated in Figure 2, and the corresponding $r_\wedge$s are presented in Table 5. Clearly by properly taking the aspect ratio into account, the correct $r_\wedge$ can always be obtained. A tendency for loss to increase with increasing aspect ratio is also seen for samples NB54 in Figure 3. Here, however, the rate of increase is slower than expected, leading to a spectacular increase in $r_\wedge$ with $a/b$ as illustrated in Figure 4. For round strands, the extracted value of $r_\wedge$ is about 6 n$\Omega$cm (only about twice the expected value of 3 n$\Omega$ based on a $RR$ (4.2K) of 500, extrapolated from the direct four point resistivity measurements discussed above) and hence could be regarded as "normal". But, as aspect ratio increases, the calculated $r_\wedge$ increases rapidly following the heuristic model based on the filamentary array geometry described above.

Values of $r_\wedge$ were also obtained for the HTSC tapes, Table 6. It is of course much more difficult experimentally to obtain a large, uniform data set for BI-type strands. However, it can be generally seen that for the round BI sample, $r_\wedge$ is close to expectation, but for the more aspected samples, with very flattened filaments, Eqn. (3) leads to a series of $r_\wedge$ values that can be an order of magnitude larger than this.

We note in passing that in MF HTSC/Ag strands the $dH/dt$ dependence of $\Delta M(H)$ is influenced by logarithmic and exponential decay of the intrafilamentary critical state and by the strand's exponential eddy current decay. In previous papers we have considered these dynamic magnetization signatures in detail and have defined two approaches to identifying the regimes of true eddy current decay[8,10]. These effects were found to be negligible[9] for these wires at 4.2 K.

## Discussion

In interpreting these results, we wish to make clear that it is not the actual matrix resistivity that becomes enhanced in these untwisted, aspected strands with coarse arrays and aspected filaments. Rather, $r_\wedge$ is taken as a useful measure of the **loss** of the strand that can be expected for a given $L$ and $dH/dt$, and this loss is suppressed because the volume through which the eddy currents are flowing are reduced by the character of the filamentary array (in conjunction with the current paths associated with the lack of twist). It would be fair to describe this as a "volume normalization effect", as long as it is recognized that this volume is an intensive property of the array and its associated current paths (for this untwisted case).

Both the (linear) proportionality of loss to $n$ or $a/b$ in aspected fine-array strands, and the increase in $r_\wedge$ for coarse-array strands are associated with the untwisted nature of the samples in this paper. That in itself is useful for interpreting the results of short sample AC loss measurements. However, we note that the expected current paths for twisted strands with resistive barriers, if we may make an analogy to superconducting Rutherford cables[6], is similar to the current paths for untwisted strands. For this reason we may expect a very significant loss suppression for these types of strands, even above that which would be expected for fine-array strands.

**Summary and Conclusions**

We have shown that for twisted MF strands, $Q_e$ is described by a treatment in which the loss expression is separable into two factors: (i) a demagnetization-associated shape-factor, $n$, equal to the strand's aspect ratio, $a/b$, and (ii) an intrastrand "materials/design relaxation factor", $t$, that embodies $r_\wedge$, $L$, and the again $a/b$ to the extent that it modifies the eddy current paths. For a coarse array, however, the loss of untwisted strands does not increase as rapidly as anticipated by this expression. This has been demonstrated by measurements both on NbTi/Cu matrix strands with low numbers of filaments as well as on Bi:2223/Ag strands with small filament numbers. The anomalously high $r_\wedge$ is attributed to the influence of the flattened filament array as well as filamentary array coarseness on the transverse current paths. This effect is expected to be much smaller for twisted strands, expect for the case of twisted strands with large interstrand barriers.

**Acknowledgements**

We thank S.X. Dou and N.V. Vo of the University of Wollongong for the HTSC samples, and E. Gregory (IGC Advanced Superconductors) for the strand NB54. The research was supported by the Electric Power Research Institute under Agreement No. WO 9006-06 and the U.S. Department of Energy, Division of High Energy Physics, under Grant No. DE-FG02-95ER40900.

# List of Tables

Table 1. Strand Material Specifications

| Strand Material | NB5000 | NB54 | BI36 | BI36R | BI19 | BI9 |
|---|---|---|---|---|---|---|
| Number of Filaments | 5355 | 54 | 36 | 36[1] | 19 | 9 |
| Matrix Material | Cu-0.6At Mn | Cu | Ag | Ag[1] | Ag | Ag |
| Strand Diameter, $d_s$, mm | 2.58 | 2.32 | 4.0 x 0.28 | 3.0 x 0.32 | 2.03 | 3.0 x 0.24 |
| Bundle Diameter | 1.69 | 0.118 (0.044) | 3.6 x 0.17 | 2.6 x 0.18 | 1.50 | 2.6 x 0.16 |
| Array Geometry | Fully Fil. | Annular Array | Fully Fil. | Fully Fil. | Fully Fil. | Fully Fil. |
| Filament Diameter, $d_f$, μm | 1.90 | 95.18 | 370 x 8 | 280 x 20 | 173.79 | 710 x 17 |
| Array Filling Factor, $I_{loc}$ | 0.675 | 0.726[2] | 0.34 | 0.34 | 0.356 | 0.34 |
| Overall Filling Factor, $I_{tot}$ | 0.402 | 0.180 | 0.218 | 0.218 | 0.194 | 0.245 |

[1]The outer shell of this sample was sterling silver
[2]Within the annular array. If we include the center in the matrix area, then $I_{loc}$ becomes 0.704.

Table 2. NbTi/CuMn 5000 Filament Samples

| Sample Name | Aspecting Method | Aspect Ratio (Strand) | Aspect Ratio (Array) | Aspect Ratio (Filament) | Strand $w$ x $t$, mm | Length, $L$, cm | Fil. Array Volume, $10^{-3}$ cm$^3$ |
|---|---|---|---|---|---|---|---|
| NB5000-1 | None | 1 | 1 | 1 | 2.58 | 1.335 | 41.6 |
| NB5000-3 | Filing | 4.13 | 3 | 1 | 2.58 x 0.625 | 1.077 | 10.35 |
| NB5000-7 | Pressing | 5.20 | 6.85 | 1-2[1] | 5.04 x .97 | 1.563 | 45.4 |
| NB5000-12 | Rolling | 10.2 | 12.35 | 3-16[1] | 5.10 x 0.50 | 1.503 | 21.60 |
| NB5000-15 | Rolling | 12.29 | 15 | -- | 6.03 x 0.49 | 1.096 | 77.14 |

[1] Irregular.

Table 3. NbTi/Cu 54 Filament Samples (Rolled)

| Sample Name | Aspect Ratio (Strand) | Aspect Ratio (Array) | Aspect Ratio (filament) | Strand $w$ x $t$, mm | Length, $L$, cm | Fil. Array Volume[1], $10^{-3}$ cm$^3$ |
|---|---|---|---|---|---|---|
| NB54-1 | 1 | 1 | 1 | 2.32 x 2.32 | 1.569 | 16.91 |
| NB54-1.5 | 1.3 | 1.53 | 1.18 | 2.51 x 1.93 | 1.5105 | 18.66 |
| NB54-2 | 1.64 | 1.97 | 1.01 | 2.76 x 1.68 | 1.561 | 18.46 |
| NB54-2.5 | 1.92 | 2.34 | 1.03 | 2.94 x 1.52 | 1.543 | 17.58 |
| NB54-3. | 2.88 | 3.34 | 1.37 | 3.46 x 1.20 | 1.527 | 16.17 |
| NB54-6 | 5.23 | 6.2 | 1.45 | 4.24 x 0.81 | 1.525 | 13.36 |
| NB54-13 | 11.4 | 13.15 | 2.08 | 5.15 x 0.45 | 1.509 | 17.82 |
| NB54-17 | 12.55 | 16.67 | 2.00 | 5.52 x 0.44 | 1.508 | 18.67 |
| NB54-22 | 23.6 | 21.33 | 2.66 | 6.61 x 0.28 | 1.493 | 21.11 |
| NB54-30 | 32.4 | 29.5 | 2.76 | 6.00 x 0.185 | 1.512 | 12.85 |

[1] In this case the volume of the center hole in the annulus is counted in this number.

Table 4. Bi:2223/Ag 36 and Filament Samples (as received)

| Sample Name | Aspect Ratio (Strand) | Aspect Ratio (Array) | Aspect Ratio (Filament) | Strand $w$ x $t$, mm | Length, $L$, cm | Fil. Array Volume, $10^{-3}$ cm$^3$ |
|---|---|---|---|---|---|---|
| BI36-21 | 14.3 | 21.2 | 46 | 4.0 x 0.28 | 1.48 | 10.9 |
| BI9-16 | 12.5 | 16.13 | 48 | 3.0 x 0.24 | 1.52 | 7.89 |
| BI36R-15 | 9.4 | 15 | 14 | 3.0 x 0.32 | 1.54 | 8.37 |
| BI36-AL-11 | 7.2 | 10.6 | 23 | 2.0 x 0.28 | 0.70 | 5.02 |
| | | | | | | |
| BI19-1 | 1 | 1 | 1[1] | 2.03 | 1.306 | 22.99 |
| BI19-7 | 5.8 | 7.45 | 10.79 | 3.02 x 0.52 | 1.330 | 11.39 |
| BI19-13 | 10.3 | 13.23 | 12.69 | 3.09 x 0.30 | 1.235 | 6.24 |
| BI19-19 | 16.15 | 18.72 | 15.2 | 3.23 x 0.20 | 1.340 | 4.72 |

[1] Irregular

Table 5. Loss-computed matrix resistivities for Nb5000/CuMn

| Sample name | Fil. Array aspect ratio, $a/b$ | EO $(dt/dH)(Q/L^2)$ s erg/Oe cm$^5$ FRV[2] | EO-derived $r_\wedge$, $10^{-7}$ $\Omega$cm | FO $(dt/dH)(Q/L^2)$ s erg/Oe cm$^5$ FRV[2] | FO-derived $r_\wedge$, $10^{-7}$ $\Omega$cm |
|---|---|---|---|---|---|
| NB5000-1 | 1 | 21.5 | 3.2 | 21.5 | 3.2 |
| NB5000-3 | 3 | 14.11[3] | 2.4 | 47.6 | 2.2 |
| NB5000-7 | 6.85 | 14.6 | 2.3 | 57.2 | 4.0 |
| NB5000-12 | 12.35 | 20.7 | 1.7 | 76.1 | 5.6 |
| NB5000-15 | 15 | -- | -- | 110 | 4.7 |

[1] Round model $\rho_\perp$ for 0.2 mm diam. strand is 3.2 x $10^{-7}$ $\Omega$cm from earlier measurements.
[2] Filamentary region volume.
[3] Filamentary region to total strand volume is 0.73 for this sample, because it is filed.

Table 6. Loss-computed transverse resistivities for BI Strands

| Sample name | FO $(dt/dH)(Q/L^2)$ s erg/Oe cm$^5$ FRV[4] | Array aspect ratio, $a/b$ | Expected[1] $r_\wedge$, n$\rho$cm | FO-derived[2] $r_\wedge$ n$\Omega$cm |
|---|---|---|---|---|
| BI36-21 | 1670 | 21.2 | 8.49 | 44[5] |
| BI9-16 | 819 | 16.1 | 8.49 | 68 |
| BI36R-15 | 595 | 15.0 | 8.49 | 87 |
| BI36-AL-11 | 426 | 10.6 | 8.49 | 86 |
| | | | | |
| BI19-1 | 276 | 1 | 8.49 | 25.0 |
| BI19-7 | 539 | 7.45 | 8.49 | 47.6 |
| BI-19-13 | 495 | 13.2 | 8.49 | 92.0 |
| BI19-19 | 540 | 18.7 | 8.49 | 119 |

[1] Expected $r_\wedge = [(1+l)/(1-l)]r_{Ag,bulk}$, where the filling factor $l = 0.34$
[2] Derived using Equation (3) with $n = a/b$
[3] Resistive sterling-silver shell
[4] Filamentary Region Volume
[5] These values differ from those in Ref 8 because here we have used filamentary array aspect ratios, rather than gross strand aspect ratios.

# List of Figures

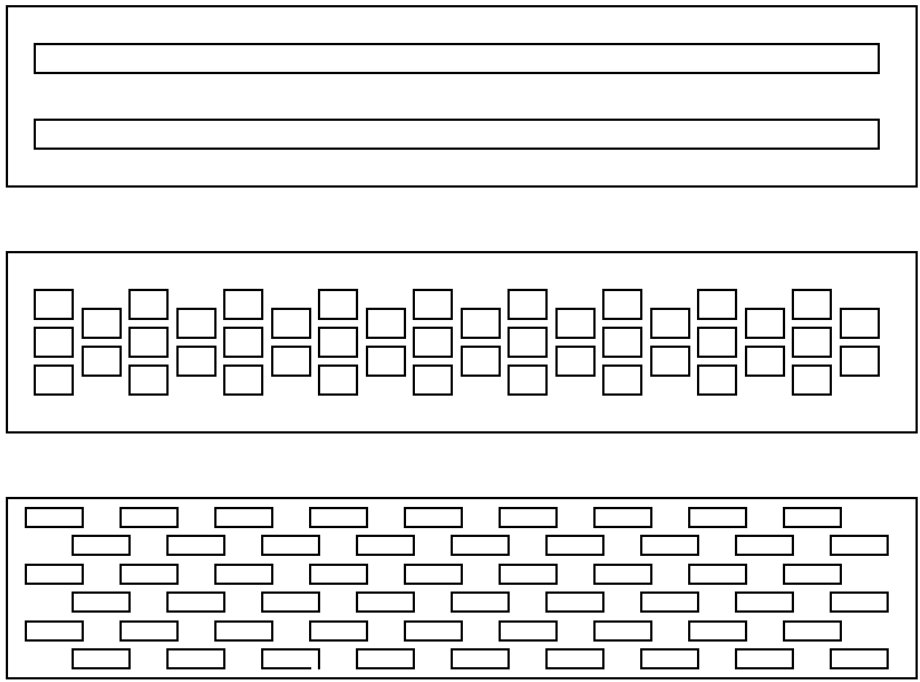

Figure 1. Various internal array geometries; (top) z-stack, (center) interpenetrating, and (bottom) non-interpenetrating.

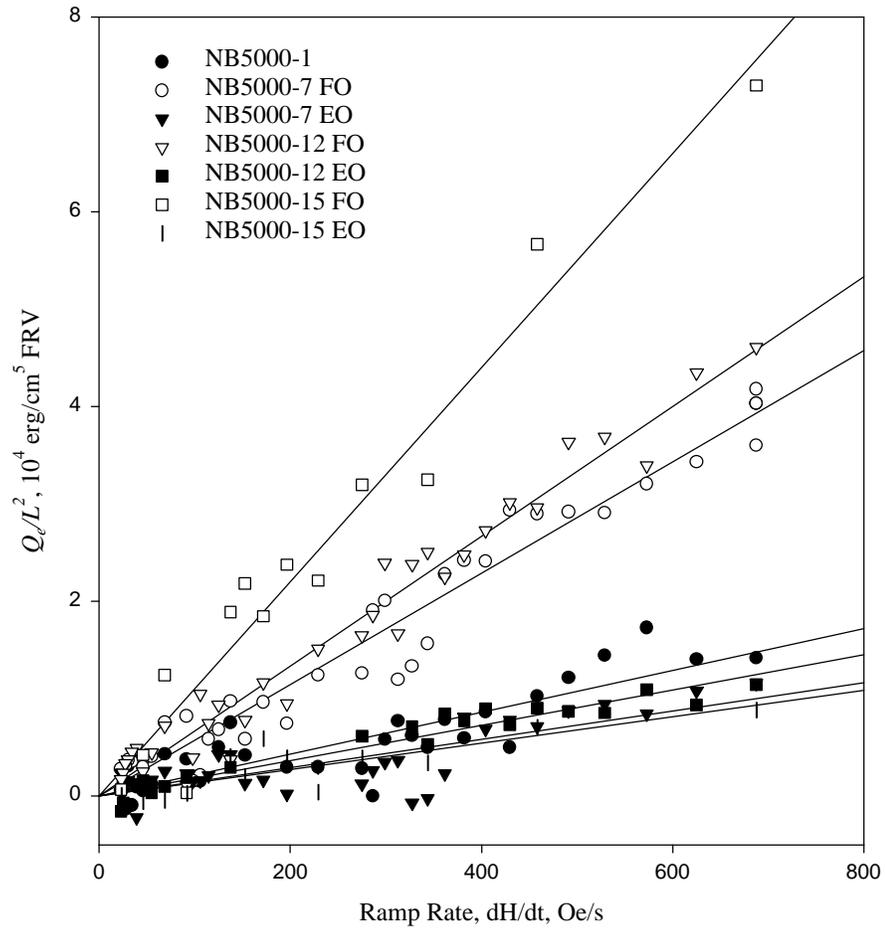

Figure 2. $Q_e/L^2$ vs $dH/dt$ for NB5000 samples.

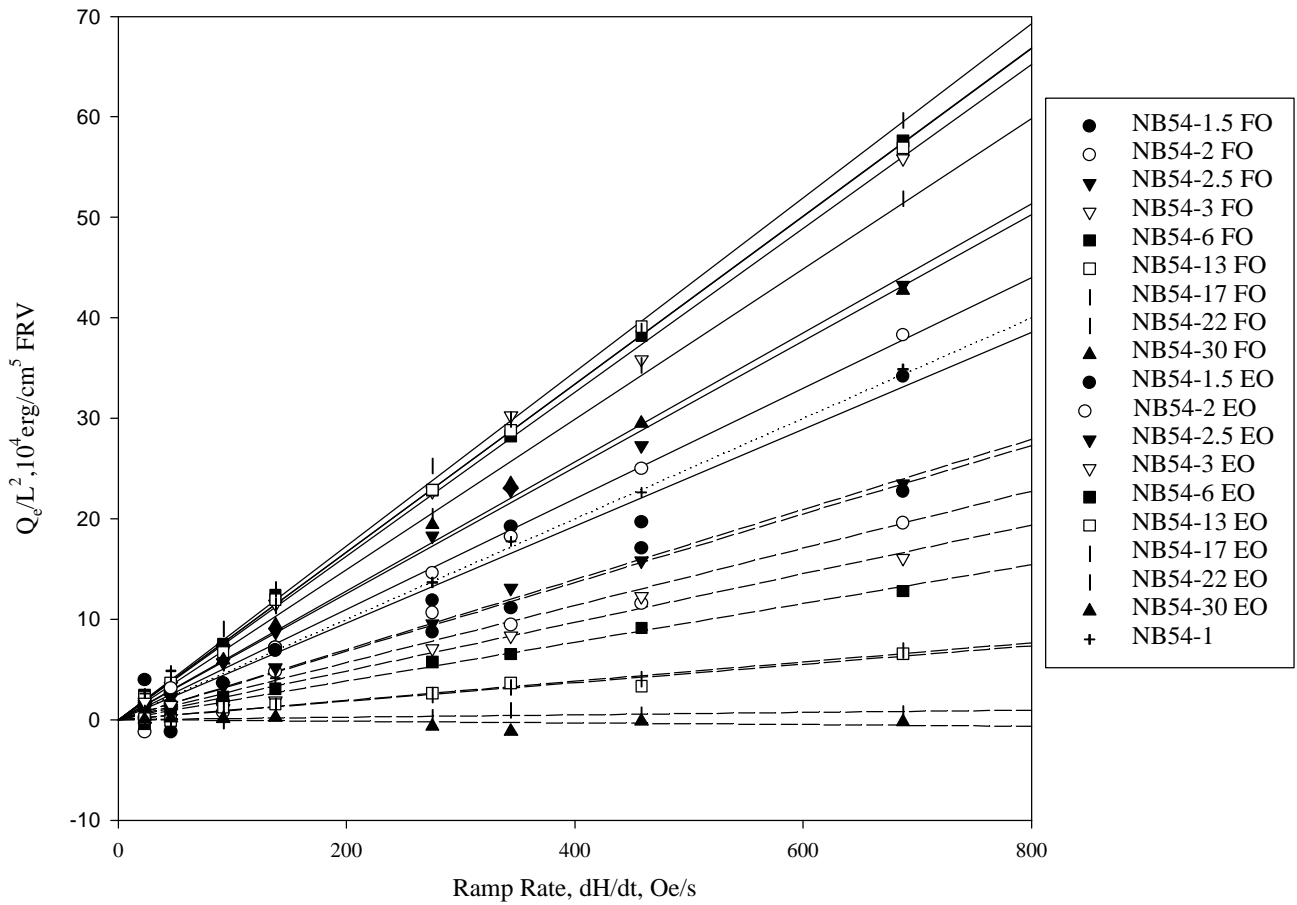

Figure 3. $Q_e/L^2$ vs $dH/dt$ for NB54 samples

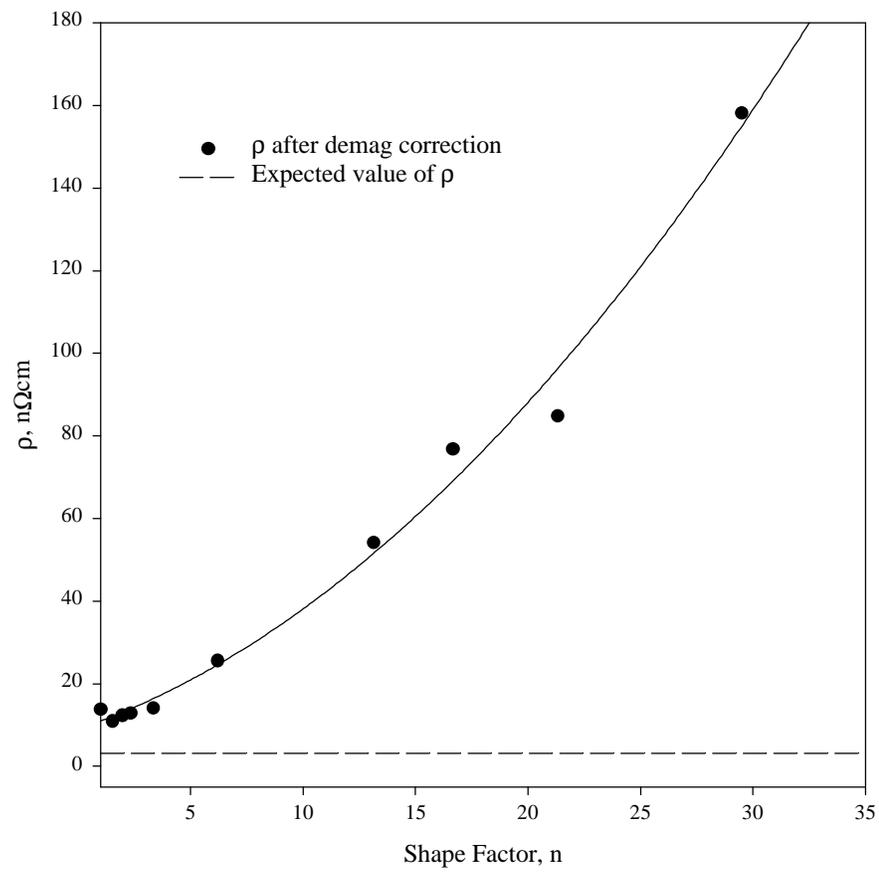

Figure 4. Resistivity vs shape factor for NB54 samples